\documentclass{article}

\usepackage{arxiv}

\usepackage[utf8]{inputenc} 
\usepackage[T1]{fontenc}    
\usepackage{amsmath,amsfonts,amssymb}
\usepackage{graphicx}
\usepackage{setspace}
\usepackage{tocloft}

\usepackage{lipsum} 
\usepackage{mathrsfs,amsmath}
\usepackage{comment}
\usepackage{scalerel}
\usepackage[overload]{empheq}
\usepackage{nicefrac,xfrac}
\usepackage{mathtools}
\usepackage{gensymb}
\usepackage{soul}
\usepackage{textgreek}

\usepackage{amsmath}
\usepackage{bm}
\usepackage{algorithm}
\usepackage{algpseudocode}
\usepackage{enumitem}
\usepackage{amsmath}
\usepackage{setspace}
\usepackage{xcolor}
\usepackage{gensymb}
\usepackage{multicol,lipsum}

\usepackage{mathtools}
\usepackage{cuted}

\usepackage{lineno}

\usepackage{authblk}

\providecommand{\vx}{\mathbf{x}}

\providecommand{\vrr}{\mathbf{r}}
\providecommand{\vu}{\mathbf{u}}

\providecommand{\vv}{\mathbf{v}}

\DeclarePairedDelimiter\abs{\lvert}{\rvert}%
\DeclarePairedDelimiter\Bigabs{\Big\lvert}{\Big\rvert}%
\DeclarePairedDelimiter\norm{\lVert}{\rVert}%

\providecommand{\tofu}{\emph{T\textsuperscript{2}oFu}}

\DeclareMathOperator*{\argmin}{arg\,min}%

\title{Tensorial tomographic Fourier Ptychography with applications to muscle tissue imaging}

\author[1]{Shiqi Xu}
\author[1,2]{Xiang Dai}
\author[3]{Paul Ritter}
\author[1,4]{Kyung Chul Lee}
\author[1]{Xi Yang}
\author[1,3]{Lucas Kreiss}
\author[1,5]{Kevin C. Zhou}
\author[1]{Kanghyun Kim}
\author[1]{Amey Chaware}
\author[6]{Jadee Neff}
\author[6]{Carolyn Glass}
\author[4]{Seung Ah Lee}
\author[3]{Oliver Friedrich}
\author[1,*]{Roarke Horstmeyer}

\affil[1]{Duke University, Durham, NC, USA 27708}
\affil[2]{UC San Diego, La Jolla, CA, USA 92093}
\affil[3]{Friedrich-Alexander University, Erlangen, Germany 91052}
\affil[4]{Yonsei University, Seoul, South Korea 03722}
\affil[5]{UC Berkeley, Berkeley, CA, USA 94720}
\affil[6]{Duke Hospital, Durham, NC, USA 27710}

\affil[*]{Corresponding author: roarke.w.horstmeyer@duke.edu}

\begin{document}
\maketitle

\begin{abstract}
We report \textit{\textbf{T}ensorial \textbf{to}mographic \textbf{F}o\textbf{u}rier Ptychography} (\tofu), a new non-scanning label-free tomographic microscopy method for simultaneous imaging of quantitative phase and anisotropic specimen information in 3D. Built upon Fourier Ptychography, a quantitative phase imaging technique, \tofu~additionally highlights the vectorial nature of light. The imaging setup consists of a standard microscope equipped with an LED matrix, a polarization generator, and a polarization-sensitive camera. Permittivity tensors of anisotropic samples are computationally recovered from polarized intensity measurements across three dimensions. We demonstrate \tofu's efficiency through volumetric reconstructions of refractive index, birefringence, and orientation for various validation samples, as well as tissue samples from muscle fibers and diseased heart tissue. Our reconstructions of muscle fibers resolve their 3D fine-filament structure and yield consistent morphological measurements compared to gold-standard second harmonic generation scanning confocal microscope images found in the literature. Additionally, we demonstrate reconstructions of a heart tissue sample that carries important polarization information for detecting cardiac amyloidosis.
\end{abstract}

\section{Introduction}
\label{sect:intro}  
Quantitative phase imaging (QPI) is a well-known label-free microscopy approach that can detect phase delay introduced by semi-transparent cells and tissue~\cite{park2018quantitative}. Due in part to its ability to provide quantitative information about primarily transparent biological specimens with low phototoxicity, QPI has become an invaluable tool in scientific and clinical studies, including for monitoring neuronal firing~\cite{ling2020high} and cancer cell line detection~\cite{el2018quantitative}, to name a few. Besides scalar phase contrast, transparent specimens also exhibit alternative and important endogenous optical contrast mechanisms, including anisotropic properties such as material birefringence and orientation. Indeed, the orientation of molecular arrangements in lipid membranes can now be quantitatively monitored with polarization-sensitive microscopes for studying multi-organelle interactive activities~\cite {zhanghao2020high,lu2020single}. There are naturally a variety of polarization-sensitive microscope arrangements, including early analog designs~\cite{schmidt1924bausteine, inoue1953polarization}, differential interference contrast (DIC) methods~\cite{nomarski1955differential}, and contemporary digital approaches that reconstruct quantitative specimen retardance and orientation~\cite{oldenbourg2013polarized, mehta2013polarized, spiesz2011quantitative}. These methods have been applied to study small model organisms~\cite{tadayon2015mantis, le2019zebrafish} and to assist clinical diagnosis~\cite{junqueira1979picrosirius,desai2010cardiac,pirnstill2015malaria, liu2019distinguishing}. 

Recently, there has been increased interest in measuring polarization-sensitive phase information from specimens. In general, polarization-sensitive quantitative phase imaging (PS-QPI) methods can be divided into two categories: those that rely on interferometric detection, and those that utilize computational phase retrieval methods. Interferometric methods (e.g., off-axis holography) can create polarization-sensitive phase images with as few as a single measurement~\cite{jiao2020real,shin2018reference,ge2021single,liu2020deep}, and can be extended to 3D with diffraction tomography approaches~\cite{van2020polarization,saba2021polarization,shin2022tomographic,taddese2023jones}. While often impressive, these methods usually require complex arrangements of coherent laser illumination and careful system alignment, which sets additional design requirements for use in clinical applications. Computational phase retrieval methods, on the other hand, rely on image reconstruction algorithms that convert multiple intensity measurements into phase-sensitive outputs~\cite{dai2022quantitative,song2020ptychography,song2021large,hur2021polarization,guo2020revealing,baroni2019joint}, and can be implemented with less expensive hardware. Due to their simple instrumentation, computational polarization microscopes have been increasingly applied to image biological samples, for instance, to study white matter tracts within whole brain slices~\cite{guo2020revealing} and to diagnose malaria from blood smears~\cite{song2021large}. Furthermore, these methods have been extended to image 3D samples, such as axons and cardiac tissue with axial scanning~\cite{yeh2021upti,xu2022tensorial}. Thick samples such as organoids and tissue slices also naturally have intriguing 3D structures that can include anisotropic material. Creating high-resolution volumetric representations of the polarization properties of these samples is essential to studying biology and pathology. While tomographic imaging methods such as confocal-based approaches have been developed in the past to image muscle tissue and neural organoids, for instance~\cite{both2004second,seo2021symmetry}, there remain relatively few microscopic techniques to jointly capture quantitative phase and anisotropy across a large three-dimensional volume at high resolution. One very recent study attempted to create 3D anisotropy maps using off-axis LED illumination but did not provide tomographic permittivity matrix reconstructions~\cite{joo2022polarization}. 

Here, we propose a non-scanning polarization-sensitive tomography method, termed \textit{\textbf{T}ensorial \textbf{to}mographic \textbf{F}o\textbf{u}rier Ptychography} (\tofu), to create quantitative volumetric permittivity matrix reconstructions without any moving parts. Our method is an extension of recently developed intensity optical diffraction tomography principles~\cite{li2019high,horstmeyer2016diffraction,pham2018versatile,lodhi2020inverse,li2022transport,chowdhury2019high} and has the potential to be extended to a video-rate system in the future~\cite{li2019high}. We image a variety of calibration targets, as well as 3D tissue and muscle fibers samples, through which we demonstrate the ability to resolve the fine filament structure and are consistent with one of the gold standards, i.e., second harmonic images, found in the literature~\cite{both2004second}. Additionally, we demonstrate reconstructions of a heart tissue sample that carries important information for detecting cardiac amyloidosis.

\section{Methods}
\begin{figure*}[!t]
\centering
\includegraphics[width=14cm]{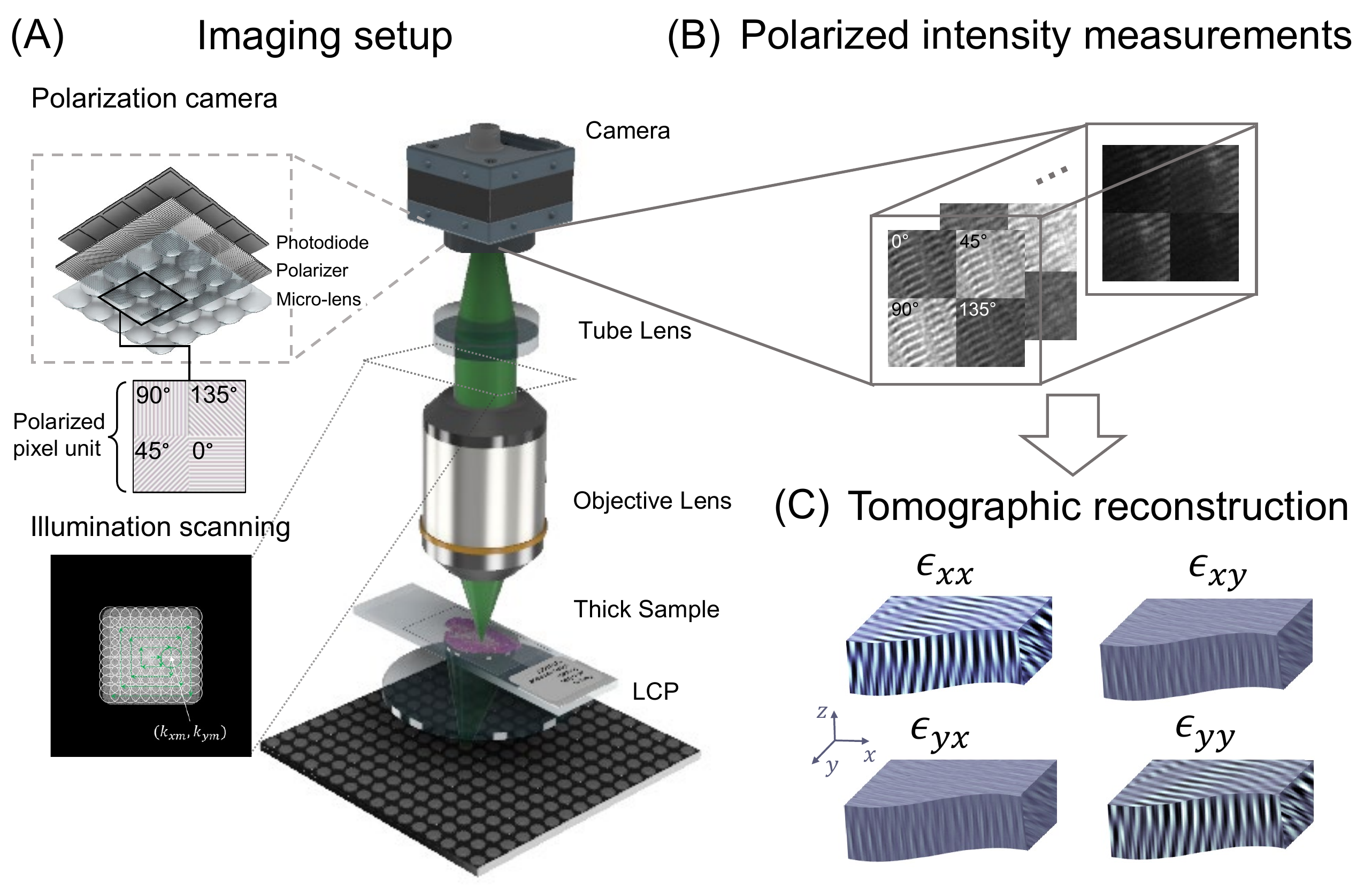}
\caption{Workflow of the proposed method. (A) An illustration of \tofu's experimental setup. Green light from an addressable LED array is circularly polarized with a left circular polarizer (LCP) to illuminate the sample. Sequential illuminations from various angles are used to scan the object in the spatial frequency domain. The sample is then imaged with an infinity-corrected optical system. The polarized light intensity at $0\degree, 45\degree, 90\degree$, and $135\degree$ are recorded with a polarization-sensitive CMOS camera as images illustrated in (B). (C) Those intensity measurements are then fused to form volumetric sample permittivity tensor reconstructions by solving the corresponding inverse problem.}
\label{fig1::overview}
\end{figure*}

\subsection{Experimental setup}
The imaging setup of \tofu\;is illustrated in Fig.~\ref{fig1::overview}(A). The illumination consists of an array of $25\times25$ addressable LEDs (\textmu Inventions Inc; Portugal) and a left-circular polarizer (CP42HE; Edmund Co., USA). $521\,\text{nm}$ wavelength light from the LED array is circularly polarized and illuminates the 3D sample from different angles. A microcontroller (ARM Cortex-M3) and a voltage level shifter (SN74AHCT) are used to turn on each small LED pixel (WS2812b-2020) sequentially. The optical field then passes through an optical system consisting of an infinity-corrected objective (0.25 NA or 0.4NA, Plan Achromatic; Olympus, JP) and a tube lens (Achromatic Doublets, 180mm focal length; Thorlabs, Inc, USA). The largest illumination NA ($NA_{illum}$) is chosen to match the native NA of the objective ($NA_{obj}$). Hence, the theoretical lateral and axial resolutions are $\delta_{x,y}=\nicefrac{\lambda}{2(NA_{obj}+NA_{illum})}$ and $\delta_z=\nicefrac{\lambda}{(2-\sqrt{1-{NA}_{obj}^2}-\sqrt{1-{NA}_{illum}^2})}$, respectively~\cite{tian20153d}. For the 0.25 NA system, $\delta_{x,y}=5.2\,\mu\mathrm{m}$ and $\delta_{z}=8.2\,\mu\mathrm{m}$. For the 0.4 NA system,  $\delta_{x,y}=3.2\,\mu\mathrm{m}$ and $\delta_{z}=3.1\,\mu\mathrm{m}$. The intensity image is captured with a polarization-sensitive CMOS camera (BFS-U3-51S5P; Teledyne FLIR LLC; USA). The polarization camera analyzes the light orientated at four different polarizations ($0\degree, 45\degree, 90\degree,$ and $135\degree$), achieved by placing $2\times2$ wire-grid polarizers between the pixel array and the microlens array, as illustrated in the m-dashed box in Fig.~\ref{fig1::overview}(A). Exemplary intensity images under illumination from different angles are shown in the box in Fig.~\ref{fig1::overview}(B).
\subsection{Principles of Tensorial tomographic Fourier Ptychography}
\subsubsection{Notation}
Here we introduce the notation we use in the rest of this article. First, we use $\vec{\cdot}$ and $\bar{\bar{\cdot}}$ symbols to denote vectors and matrices for variables, respectively. By default, all the vectors in this article are column vectors. Second, we utilize the \textit{Mathematical Script} font enclosed in curly braces to denote spatial operators. As an example, one frequently used operator is $\mathcal{F}\{\cdot\}$, which indicates the Fourier transform in space. Next, \textbf{bold letters} in lowercase represents support vectors in either frequency ($\vu$) or space ($\vrr$). Finally, we utilize the $\tilde{\cdot}$ symbol to denote the frequency-domain counterparts of variables previously defined in the space domain, such as $\tilde{I}(\vu) = \mathcal{F}\{I(\vrr)\}$.

\subsubsection{Vectorial light propagation}
\label{sec::physics}
The optical properties of a 3D sample can be described by its $3\times3$ permittivity matrix~\cite{born2013principles} 
\begin{equation}
\bar{\bar{\epsilon}}=
    \begin{bmatrix}
\epsilon_{xx}(\vrr) & \epsilon_{xy}(\vrr) & \epsilon_{xz}(\vrr)\\
\epsilon_{yx}(\vrr) & \epsilon_{yy}(\vrr) &
\epsilon_{yz}(\vrr)\\
\epsilon_{zx}(\vrr) & \epsilon_{zy}(\vrr) &
\epsilon_{zz}(\vrr)
\end{bmatrix},
\end{equation}
where $\vrr=(x, y, z)$ is the voxel position in space. In the scope of this work, we adopt the first Born approximation~\cite{yeh2021upti,saba2021polarization}, and the scattered vectorial electric field $\vec{E}^s$ and illumination $\vec{E}^0 $ is related as
\begin{equation}
     \vec{E}^s(\vrr)=
     \iiint\bar{\bar{G}}(\vrr-\vrr')\bar{\bar{V}}(\vrr')
     \vec{E}^0(\vrr')d\vrr',
\end{equation}
where $\bar{\bar{V}}(\vrr') = \bar{\bar{\epsilon}} - \bar{\bar{\epsilon}}_0$ is the sample scattering potential tensor with $\bar{\bar{\epsilon}}_0$ as the permitivity tensor of the background medium. $\bar{\bar{G}}(\vrr)$ is the dyadic Green's tensor~\cite{yaghjian1978direct}. In addition, we use illuminations with relatively small incident angles, which have weak polarization along the z-axis. Therefore, we mainly consider the transverse polarization of the electrical field, $\vec{E}_\perp=(E_x, E_y)^T$, which will be discussed in more detail in the next subsection. The intensities of the vectorial electric field $\vec{E}_\perp$ at different orientations are then analyzed and recorded by a polarization-sensitive optical imaging system, which we model as a $2\times2$ pupil matrix at each spatial frequency $\vu$~\cite{dai2022quantitative}. Therefore, the measured intensity under plane wave illumination with angle $\vu'$ analyzed by a polarizer with Jones vector $\vec{a}$ is
\begin{equation}
        I^{l}(\vrr,\vu')=\Bigabs{{\vec{a}_l}^T\mathcal{F}_{2d}^{-1}\Big\{\bar{\bar{P}}(\vu)\mathcal{F}_{2d}\big\{\vec{E}_{\perp}^s(\vrr,\vu')+\vec{E}_{\perp}^0(\vrr,\vu')\big\}\Big\}}^2.
\label{eq:forward_general}
\end{equation}
For a linear polarization analyzer oriented at $\alpha$, $\vec{a}=[\cos{\alpha},\sin{\alpha}]^T$~\cite{ferrand2015ptychography,dai2022quantitative}. The intuition behind this is that light can only oscillate in the same direction as the linear polarizer~\cite{dai2022quantitative}.

\subsubsection{Forward model and inverse problem}
For our initial demonstration of \tofu, we modify the model discussed in Section \ref{sec::physics} by making certain approximations. The objective of this is to craft a meaningful forward model that accurately describes our experimental measurements, whose inverse problem is less ill-posed. First, following Saba \textit{et al.} and other closely relevant works in literature~\cite{saba2021polarization,song2020ptychography,song2021large,dai2022quantitative}, we begin with a paraxial approximation, which assumes weak polarization along the optical axis of the illumination and negligible interaction between traverse and axial polarization from the sample. With this, we can simplify the $3\times3$ permittivity matrix to a $2\times2$ matrix, denoted as 
\begin{equation}
\bar{\bar{\epsilon}}=
    \begin{bmatrix}
\epsilon_{xx}(\vrr) & \epsilon_{xy}(\vrr)\\
\epsilon_{yx}(\vrr) & \epsilon_{yy}(\vrr)
\end{bmatrix}.
\label{eq:permit_define}
\end{equation}
While this approximation may not always be accurate for every anisotropic sample when illuminated at high angles, it is accurate under up to a $25\degree$ oblique illumination, based on a study using finite element analysis~\cite{saba2021polarization}. Additionally, we assume that the background media is isotropic and uniform (i.e., not spatially dependent) with a diagonal permittivity tensor $\bar{\bar{\epsilon}}_0=\epsilon_0\mathbb{I}$. This simplifies Green's tensor to a diagonal matrix with the same component for each polarization~\cite{saba2021polarization},
\begin{equation}
    \bar{\bar{G}}(\vrr,\vrr')=\begin{bmatrix}
      G(\vrr,\vrr') & 0  \\
      0 & G(\vrr,\vrr')
     \end{bmatrix}.
\end{equation}
$G(\vrr,\vrr')=G(r)$, where $r=\abs{\vrr-\vrr'}$, is the scalar Green's function that has a Weyl expansion ~\cite{born2013principles,weyl1919new}
\begin{equation}
    \frac{e^{jk_0 r}}{r}=\frac{1}{j2\pi}\int d\vu\frac{e^{-j(\vu\cdot\vx+\eta\abs{z})}}{\eta},
\end{equation}
with wavenumber vectors in lateral ($\vu=(k_x,k_y)$) and axial ($\eta=\sqrt{k_0^2-\abs{\vu}^2}$) directions. $k_0$ is the wavenumber of the isotropic background medium. This expansion is generally easier to work with when we prefer to represent the object in both lateral frequency and axial space domain. 

In addition, we make the assumption that the sample being imaged is homogeneous, which is usually assumed for many types of crystals and biological samples \cite{chipman2018polarized,mehta2013polarized}. The permittivity matrix is then symmetric and can be decomposed into~\cite{chipman2018polarized}:
\begin{equation}
\label{rot_mtx}
\bar{\bar{\epsilon}}=
    \begin{bmatrix}
\cos\theta & \sin\theta\\
-\sin\theta & \cos\theta
\end{bmatrix} \begin{bmatrix}
\epsilon_e & 0\\
0 & \epsilon_o
\end{bmatrix} \begin{bmatrix}
\cos\theta & -\sin\theta\\
\sin\theta & \cos\theta
\end{bmatrix}.
\end{equation}
The variables $\epsilon_o$ and $\epsilon_e$ represent the permittivity values along the ordinary and extraordinary axes, respectively. The parameter $\theta$ is the angle between the principle axis and the extraordinary axis (also known as the slow axis). Following the convention introduced by an early literature~\cite{jones1941new}, we rename elements in Eq.\ref{eq:permit_define} as
\begin{subequations}
\label{eq:def_eps}
    \begin{empheq}[left={\empheqlbrace\,}]{align}
     \epsilon_1 &= \epsilon_{xx}\\
     \epsilon_2 &= \epsilon_{yy}\\
     \epsilon_3 &= \epsilon_{xy}=\epsilon_{yx}.
    \end{empheq}
\end{subequations}
Since we use left circularly polarized illumination, the transverse scattering potential becomes
\begin{equation}
     \begin{bmatrix}
V_1 & V_3\\
V_3 & V_2
\end{bmatrix}\begin{bmatrix}1\\j\end{bmatrix}=\begin{bmatrix}V_1+jV_3\\V_3+jV_2\end{bmatrix},
\end{equation}
where
\begin{subequations}
\label{eq:def_v}
    \begin{empheq}[left={\empheqlbrace\,}]{align}
     V_1 &= 4\pi{k}_0^2(\epsilon_1-\epsilon_0)\\
     V_2 &= 4\pi{k}_0^2(\epsilon_2-\epsilon_0)\\
     V_3 &= 4\pi{k}_0^2\epsilon_3.
    \end{empheq}
\end{subequations}

Finally, we refer to Ling et. al.~\cite{ling2018high} for two further approximations of the illumination and scattering processes. Our first assumption is that the illumination from each LED at the sample plane is a plane wave, which is commonly used in Fourier ptychographic computational microscopy~\cite{konda2020fourier}. The second approximation involves utilizing a weak object assumption that ignores the second-order scattering term~\cite{li2019high,chen20163d,ayoub20213d}. Further, we disregard pupil aberration and model it as a low-pass filter $P(\vu)$ with a cutoff frequency based on the numerical aperture of the objective lens~\cite{li2019high}. Jones matrix has been demonstrated feasible in previous literature for the correction of anisotropic aberration by jointly reconstructing the pupil~\cite{dai2022quantitative,baroni2019joint}, which is planned for future research. Based on these approximations, the forward model can be expressed as~\cite{li2019high}
\begin{align}
\begin{split}
     \tilde{I}^{l,m}(\vu,z=0)\approx\tilde{I}_0^{l,m}(\vu,z=0)+\int\Bigr[{H}_{Re}^m&(\vu,z)\cdot\tilde{V}_{Re}^l(\vu,z)\\&+{H}_{Im}^m(\vu,z)\cdot\tilde{V}_{Im}^l(\vu,z)\Bigl]\, dz,
\end{split}
\end{align}
where $\tilde{I}^{l,m}(\vu,z)$ and $\tilde{I}_0^{l,m}(\vu,z)$ are 2D Fourier transform of the measurement and DC term from $m^{th}$ LED illumination analyzed by $l^{th}$ polarizer, respectively.

\begin{equation}
    {\small\begin{split}
    H_{Re}^m(\vu,z)=\frac{jk^2}{2}S(\vu_m)\biggl\{P^*(-\vu_m)\frac{e^{-j[\eta_i+\eta(\vu-\vu_m)]z}}{\eta(\vu-\vu_m)}&P(\vu-\vu_m)\\&-P(-\vu_m)\frac{e^{j[\eta_i+\eta(\vu+\vu_m)]z}}{\eta(\vu+\vu_m)}P(-\vu-\vu_m)\biggr\}
    \end{split}}
\end{equation}
and 
\begin{equation}
    {\small\begin{split}
    H_{Im}^m(\vu,z)=\frac{-k^2}{2}S(\vu_m)\biggl\{P^*(-\vu_m)\frac{e^{-j[\eta_i+\eta(\vu-\vu_m)]z}}{\eta(\vu-\vu_m)}&P(\vu-\vu_m)\\&+P(-\vu_m)\frac{e^{j[\eta_i+\eta(\vu+\vu_m)]z}}{\eta(\vu+\vu_m)}P(-\vu-\vu_m)\biggr\}
    \end{split}}
\end{equation}
are the diffractive transfer functions in frequency and space for the real and imaginary part of the scattering potential under $m^{th}$ LED illumination with shape ${S}^m(\vu')$~\cite{chen20163d,streibl1985three}. Since each LED has a very small die area ($<170$ \textmu{m} in diameter), we represent it as a delta function. $l\in\{0\degree,45\degree,90\degree,135\degree\}$. $\tilde{V}_{Re}^l(\vu,z)$ and $\tilde{V}_{Im}^l(\vu,z)$ are 2D Fourier transforms of ${V}_{Re}^l(\vrr)$ and ${V}_{Im}^l(\vrr)$ along lateral directions at depth $z$, respectively, and are related to Eq. \ref{eq:def_v} via
\begin{subequations}
\label{eq:def_vdegree}
    {\begin{empheq}[left={\empheqlbrace\,}]{align}
    &{V}_{Re}^{0\degree}(\vrr) + j{V}_{Im}^{0\degree}(\vrr) =  V_1(\vrr)+jV_3(\vrr)\\
     \begin{split} &{V}_{Re}^{45\degree}(\vrr) + j{V}_{Im}^{45\degree}(\vrr) = 
         \frac{\sqrt{2}}{2}(V_1(\vrr)+V_3(\vrr))\\&\quad\quad\quad\quad\quad\quad\quad\quad\quad\quad+j\frac{\sqrt{2}}{2}(V_2(\vrr)+V_3(\vrr))
     \end{split}\\
     &{V}_{Re}^{90\degree}(\vrr) + j{V}_{Im}^{90\degree}(\vrr) = V_3(\vrr)+jV_2(\vrr)\\
          \begin{split} &{V}_{Re}^{135\degree}(\vrr) + j{V}_{Im}^{135\degree}(\vrr)=  \frac{\sqrt{2}}{2}(-V_1(\vrr)+V_3(\vrr))\\&\quad\quad\quad\quad\quad\quad\quad\quad\quad\quad+j\frac{\sqrt{2}}{2}(V_2(\vrr)-V_3(\vrr)),\end{split}
    \end{empheq}}
\end{subequations}
which are the scattering potential components corresponding to each analyzer angle. We want to point out that $V_{1,2,3}(\vrr)$ are complex variables that may have imaginary parts; hence, the above equation does not imply $ \tilde{V}_{Re}(\vrr)=V_1(\vrr)$, for example. 

For concise expression, we define a new variable $\mathbf{v}\in\mathbb{C}^{N\times M\times T\times3}$ representing all the potentials $V_1(\vrr), V_2(\vrr)$, and  $V_3(\vrr)$, where $N,M,T$ are width, height, and depth of the 3D sample, respectively. Further, we introduce the operator $\mathcal{A}^{l,m}(\cdot)$ as the forward model for the sample illuminated by the $m^{th}$ LED and analyzed by the $l^{th}$ linear polarizer. To reconstruct the permittivity matrix, we formulate the inverse problem as
\begin{equation}
    \mathbf{v} = \argmin_{\mathbf{v}}\mathcal{L}(\mathbf{v}),
\end{equation}
with the loss function
\begin{equation}
\label{eq:loss_fuc_inverse}
    \mathcal{L}(\mathbf{v}) = \sum_l\sum_m\norm{\mathcal{A}^{l,m}(\mathbf{v})-{\tilde{I}}^{l,m}(\vu,z)}_2^2 + \gamma \text{tv}(\vv).
\end{equation}
tv$(\cdot)$ is the isotropic total variation operator. $\gamma$ is a regularization parameter empirically set to be $1\times10^{-6}$ for all experiments. The forward model is implemented in Pytorch and the loss function is optimized using a stochastic gradient descent method with Nesterov momentum acceleration~\cite{paszke2019pytorch,nesterov1983method}. 

Subsequently, we extract polarization properties of interest, such as orientation and birefringence from \tofu\; reconstructions $\epsilon_{1,2,3}$. The permittivity of ordinary and extraordinary axes can be computed as
\begin{subequations}
\label{eq:extra_ord}
\begin{empheq}[left={\empheqlbrace\,}]{align}
    \epsilon_0 &= \bar{\epsilon}-\nicefrac{1}{2}\Delta\epsilon \\
        \epsilon_e &= \bar{\epsilon}+\nicefrac{1}{2}\Delta\epsilon,
        \end{empheq}
\end{subequations}
and the refractive index along the ordinary and extraordinary axes $n_{o,e}=\sqrt{\epsilon_{o,e}}$ can be further derived, along with the averaged refractive index ($\bar{n}=n_o+n_e$) and birefringence ($\Delta{n}=n_e-n_o$). Moreover, the orientation to the slow axis is computed as
\begin{equation}
\theta=\begin{cases}
		  \frac{1}{2}\arctan\nicefrac{2\epsilon_3}{\epsilon_1-\epsilon_2}, \;  &\text{if}\; \epsilon_1-\epsilon_2>0\\
           \frac{1}{2}\arctan\nicefrac{2\epsilon_3}{\epsilon_1-\epsilon_2}+\frac{\pi}{2}, \; &\text{otherwise.}
		 \end{cases}
\label{eq:extra_ori}
\end{equation}


\section{Experimental Results}
\begin{figure}[!t]
\centering
\includegraphics[width=12cm]{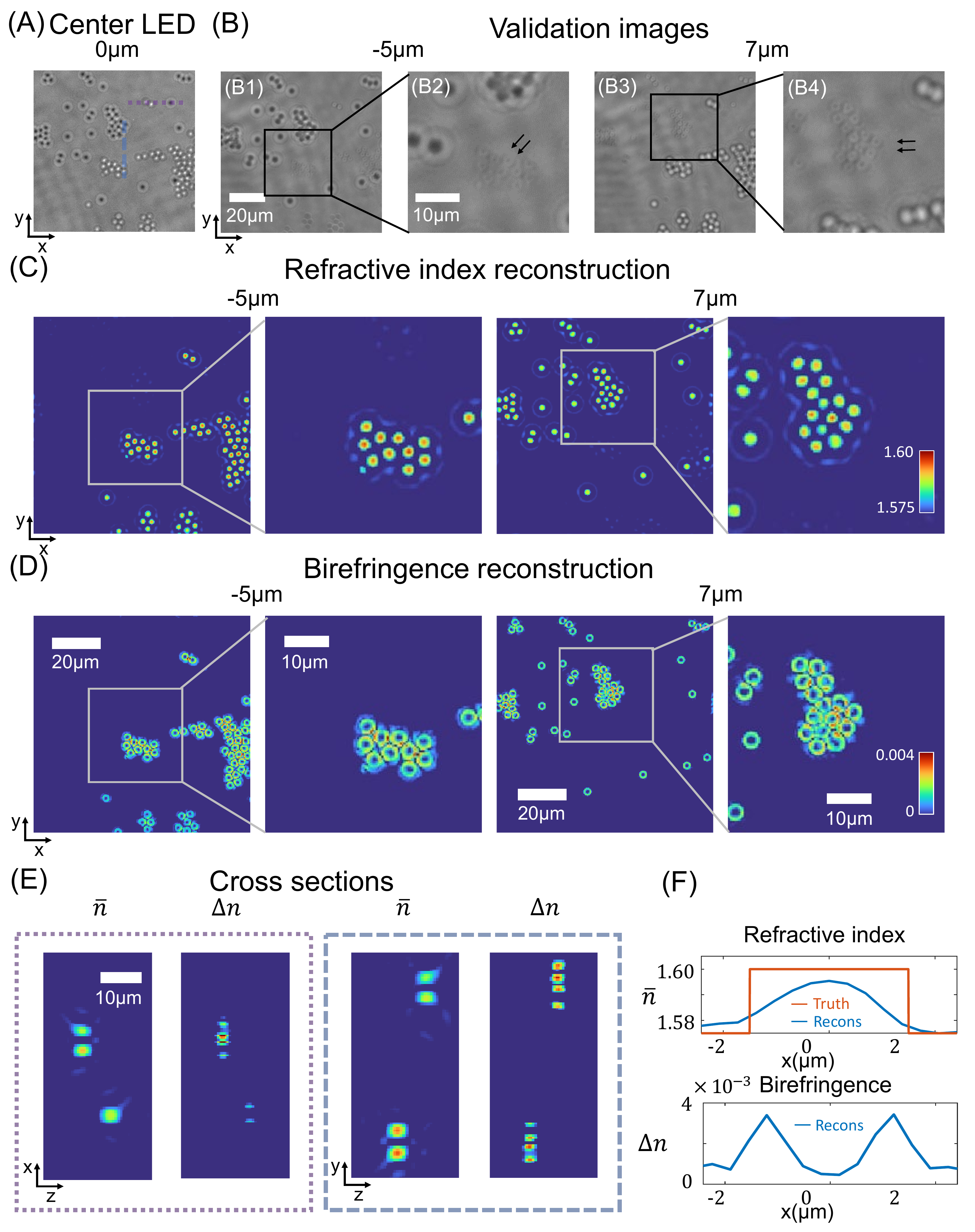}
\caption{Reconstruction results for polystyrene microspheres. (A-B) Intensity images of the sample illuminated with center LED. (A) The data is captured when focused at z=0 \textmu{m}. (B) Images were captured when focused at -5 \textmu{m} and 7 \textmu{m} by mechanically moving the sample. These images serve as a reference to be compared with the reconstruction. (C)-(E) plots the reconstructed refractive index and birefringence. All columns in (C) and (D) share the same scale bar, all the reflective index reconstructions (C), as well as all birefringence reconstructions (D) and all cross-sections (E) share the same colorbar. (E) displays cross-sections of the reconstruction in places color labeled in (A). (F) plots the profile of reconstruction averaged over 10 microspheres.}
\label{fig2::beads}
\end{figure}
\begin{figure}[!t]
\centering
\includegraphics[width=8cm]{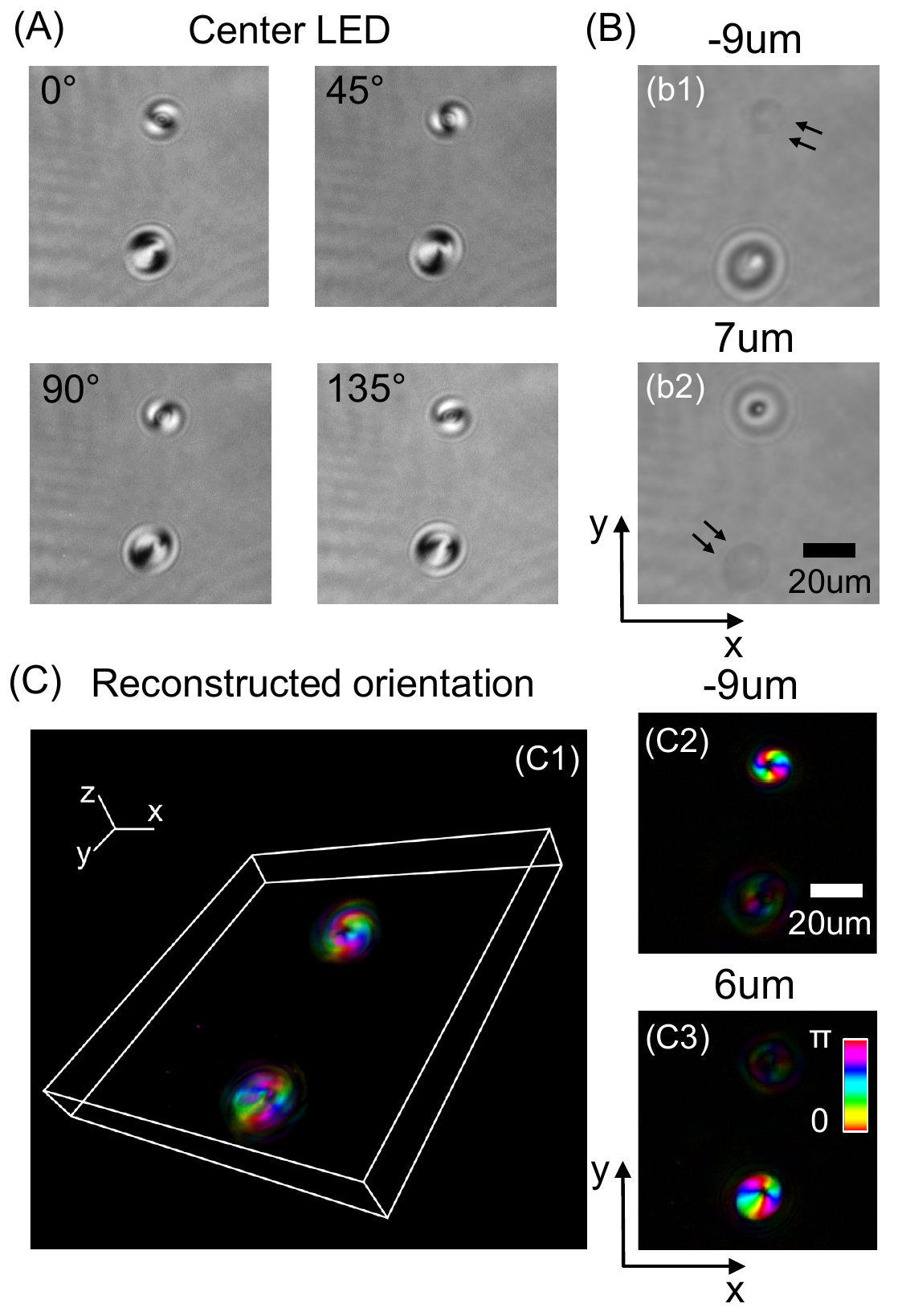}
\caption{Reconstruction results of potato starch grain. (A) Polarized intensity measurements from the center LED. (B) Images taken when focused at -9\,\textmu{m} and 7\,\textmu{m} serve as a reference for validating the reconstruction. (C) Tomographic reconstruction of the sample. The orientation is coded in color, while the birefringence is represented as brightness. All the images in (C) share the same colorbar.}
\label{fig3::potato}
\end{figure}

To validate the performance of the proposed method, we first show results from a variety of calibration targets. Due to the lack of commercially available high-resolution 3D polarization phantom, we follow previous works to validate different aspects separately~\cite{yeh2021upti,song2021large,bai2020pathological,dai2022quantitative}. We then show reconstructions of a single fixed muscle fiber. Finally, we show reconstructions of a tissue section sample from a human heart biopsy that is predictive for lethal cardiac amyloidosis. All the images presented here are captured and then reconstructed using a $20\times$, 0.4 NA system, except for the cardiac tissue, which is recorded and processed with a $10\times$, 0.25 NA system.

\subsection{Tomographic reconstruction}
In this subsection, we demonstrate tomographic reconstructions of averaged refractive index $\bar{n}$ and birefringence $\Delta n$ from isotropic and anisotropy calibration samples. These results are acquired with a $20\times$ objective ($0.4$NA, Olympus Corporation, Japan.) imaging system. Figure~\ref{fig2::beads} (A) shows the image captured with center LED illumination when the imaging system is focused at the middle of a polystyrene microsphere sample. The sample is made of two layers of $3$\textmu{m}-diameter microspheres immersed in $n=1.575$ oil. (B) shows images taken with center LED illumination when focused at different depths. The arrows in (B) highlight the microspheres that are in focus, suggesting two layers of microspheres are placed at $-7$\textmu{m} and $5$\textmu{m} planes. Note that since the refractive index of polystyrene ($n=1.60$ @ $520$nm) is very close to the background medium, the contrast of in-focus microspheres is very low. (C)-(D) shows the reconstructed refractive index and birefringence at two different depths. The accuracy of the tomographic depth reconstruction is validated with images displayed in (B). Since polystyrene is an isotropic material, the reconstruction shows no anisotropy properties except on the edge, which agrees with the well-recognized edge birefringence phenomena\cite{oldenbourg1991analysis} reported in prior literature\cite{yeh2021upti,dai2022quantitative}. (E) shows the cross-sections of the reconstructed reflective index and birefringence. (F) plots the profile of the reconstructed birefringence and reflective index averaged across 10 microspheres. 

In addition, we show reconstructions of an anisotropic potato starch sample in Fig.~\ref{fig3::potato}. The sample consists of two potato starch grains immersed in $n=1.515$ oil at different depths. Fig.~\ref{fig3::potato}(A) depicts captured intensity images illuminated with the center LED at four different polarizations. We can see the spiral patterns imaged with different polarization differs from each other noticeably. Similarly, Fig.~\ref{fig3::potato}(B) shows images when the system is focused at different depths of the sample. The black arrows point at grains that are in focus, suggesting the two potato starch grains are suspended at $-9\,$\textmu{m} and $6\,$\textmu{m}. Fig.~\ref{fig3::potato}(C) shows the reconstructed orientation and birefringence. To best visualize the results, we follow the convention~\cite{liu2020deep,song2021large,dai2022quantitative} to display this multidimensional data using an HSV colormap, where saturation is set to one, value is associated with birefringence, and orientation is coded in hue. The reconstructed structures agree with starch grain reconstructions reported in previous holography-based literatures~\cite{saba2021polarization,taddese2023jones}. 
\begin{figure*}[!t]
\centering
\includegraphics[width=13cm]{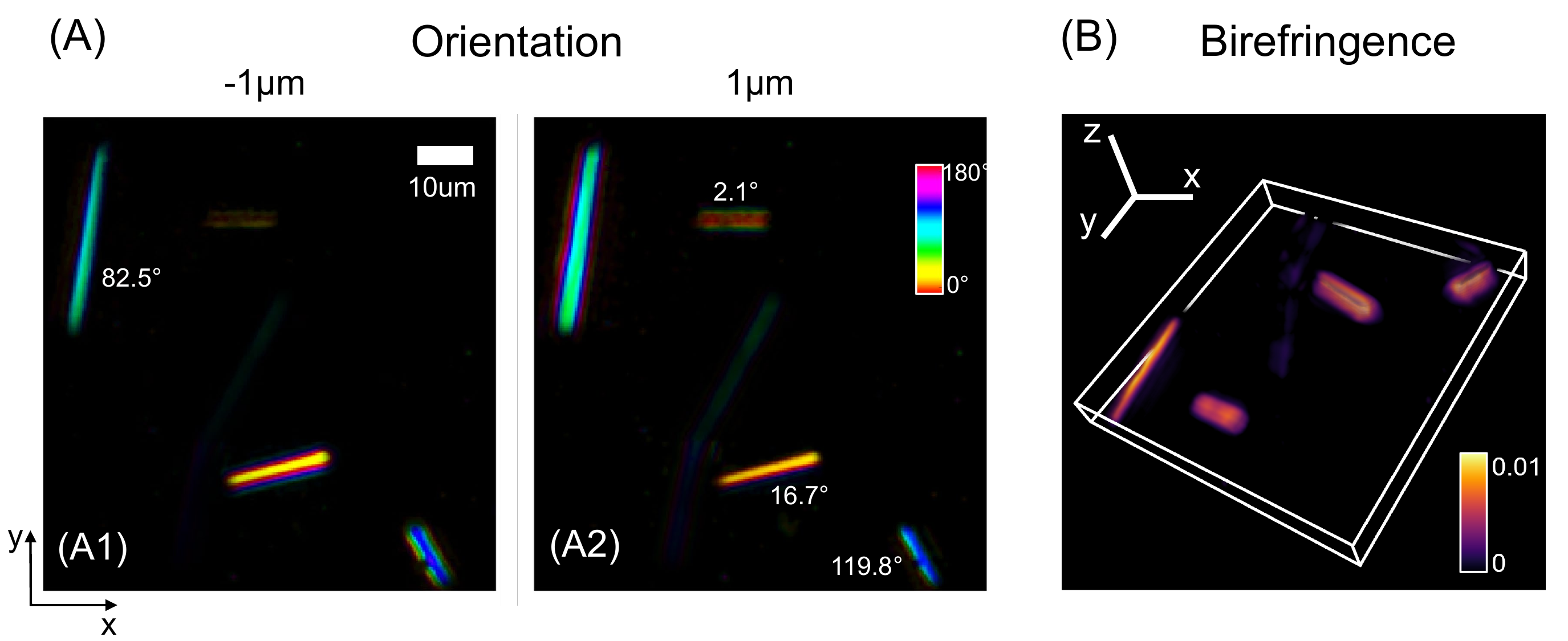}
\caption{Reconstructions of monosodium urate (MSU) crystals. (A) shows the reconstructed polarization orientation of the crystals at two different depths, labeled with their respective structural directions. (B) renders the birefringence reconstruction of the sample in 3D.}
\label{fig4::msu}
\end{figure*}

\subsection{Orientation measurement}
To verify the accuracy of the reconstructed orientation, we test our method on a sample made with monosodium urate (MSU). MSU are needle-shaped crystals precipitated from uric acid that could trigger robust inflammation such as acute arthritis and other immune activations that cause severe pain in patients~\cite{shi2010monosodium}. 
Figure~\ref{fig4::msu}(A) displays the reconstructed birefringence and orientation at two slightly different depths. These results were also obtained with a $20\times$ objective imaging system and suggest that the reconstructed orientation values of line-shaped MSU crystals follow the structural direction (labeled next to each MSU crystal). This is in agreement with reconstruction results from LC-PolScope-based methods found in the literature~\cite{song2021large,bai2020pathological}. Additionally, Figure~\ref{fig4::msu}(B) presents a 3D rendering of the reconstructed birefringence.
\subsection{Muscle fiber assessment}
\begin{figure*}[!t]
\centering
\includegraphics[width=14cm]{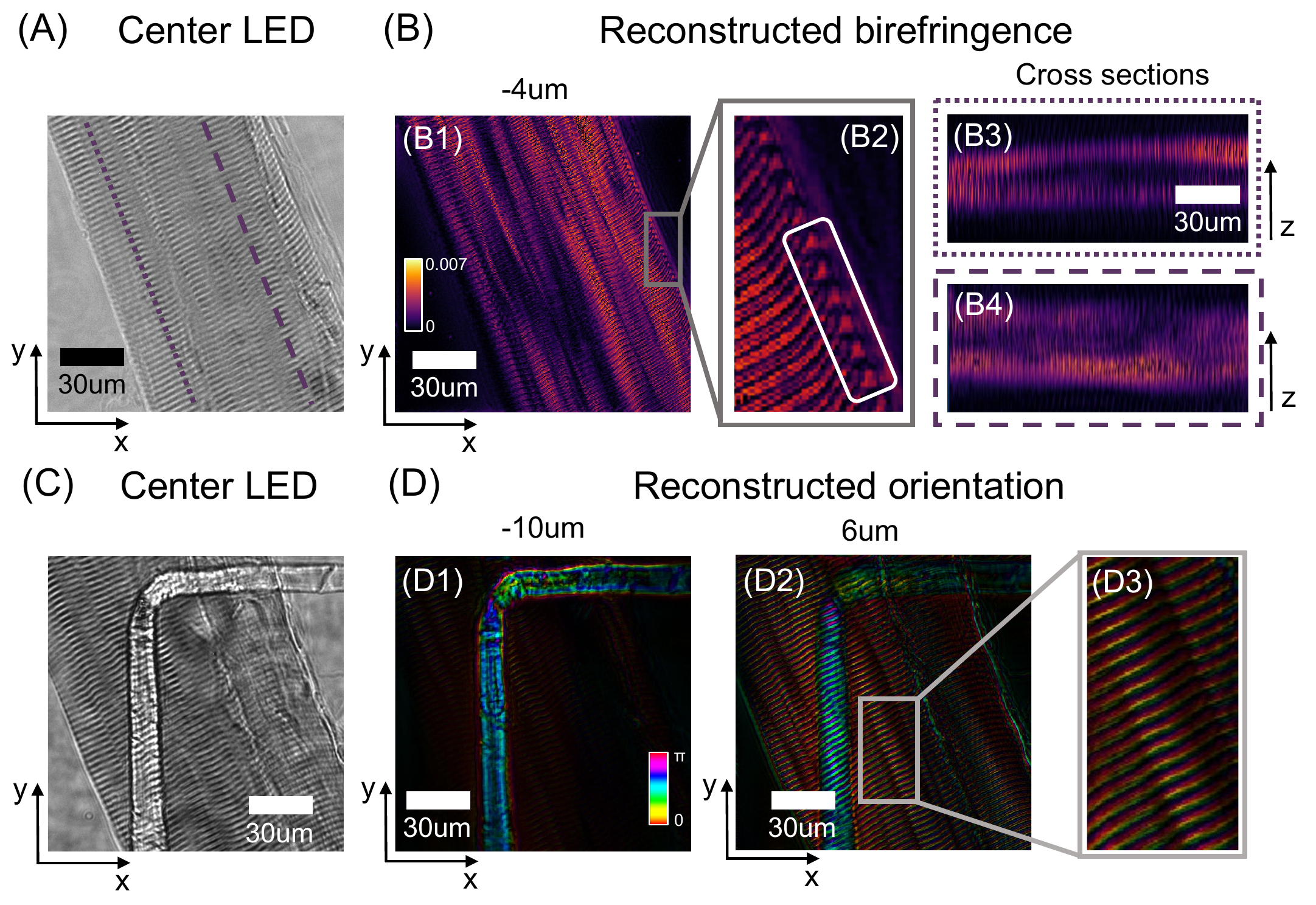}
\caption{Reconstructions of a muscle fiber. (A) shows the image of a muscle fiber with the center LED illumination. The imaging system is focused in the middle of the muscle fiber. (B) plots the reconstructed birefringence. The zoom-in region resembles reconstruction results reported in \cite{both2004second} using a second harmonic generation imaging approach. (c) shows the image of the same muscle fiber at a different region, where a non-muscle fiber with a ninety-degree bend is placed below the muscle fiber (see -10\textmu{m} in D). The imaging system is focused between this and the muscle fiber. (D) shows the reconstructed orientation at different depths, with a zoom-in showing the fine sarcomere structure of muscle tissue.}
\label{fig5::muscle}
\end{figure*}
High contrast and high-resolution structural imaging of intrinsic signals in muscle skeletal fibers is an important task for the rapid detection of changes in myofibrillar organization that can lead to skeletal myopathies. Currently, 3D muscle tissue is typically imaged by complex and expensive systems, such as second harmonic generation (SHG) microscopy. SHG exploits the contrast of polarization properties from the non-linear susceptibility in molecules like myosin, by using high-power, ultra-short pulsed lasers in a point scanning configuration~\cite{both2004second}. Here we show 3D~\tofu\, 
 reconstruction results of an isolated healthy muscle fiber using our inexpensive, LED-based and scanning-free system. Figure~\ref{fig5::muscle}(A) shows an image of a muscle fiber captured with center LED illumination. Figure~\ref{fig5::muscle}(B) shows a volumetric reconstruction of the muscle fiber. The cross sections of regions highlighted in Fig.~\ref{fig5::muscle}(A) are depicted. The zoom-in region of the image in x-y resembles reconstructions reported in ~\cite{both2004second}. Figure~\ref{fig5::muscle}(C) shows an image of the same muscle fiber from a different field of view, where a non-muscle fiber with a ninety-degree bend is placed below the muscle fiber. Figure~\ref{fig5::muscle}(D) depicts reconstructed orientation and birefringence at two different depths, showing the regular pattern of a healthy muscle fiber. The zoom-in region highlights the muscle grains with consistent orientations, in agreement with the results reported by Both et al.~\cite{both2004second}. The reconstruction shows a change in orientation at the bend of the non-muscle fiber (D1), while the orientation of the muscle fiber remains constant. Furthermore, the spatial resolution of the reconstruction is sufficient to clearly resolve the muscle filaments (D3).

\subsection{Imaging cardiac amyloidosis}
\begin{figure}[!t]
\centering
\includegraphics[width=8cm]{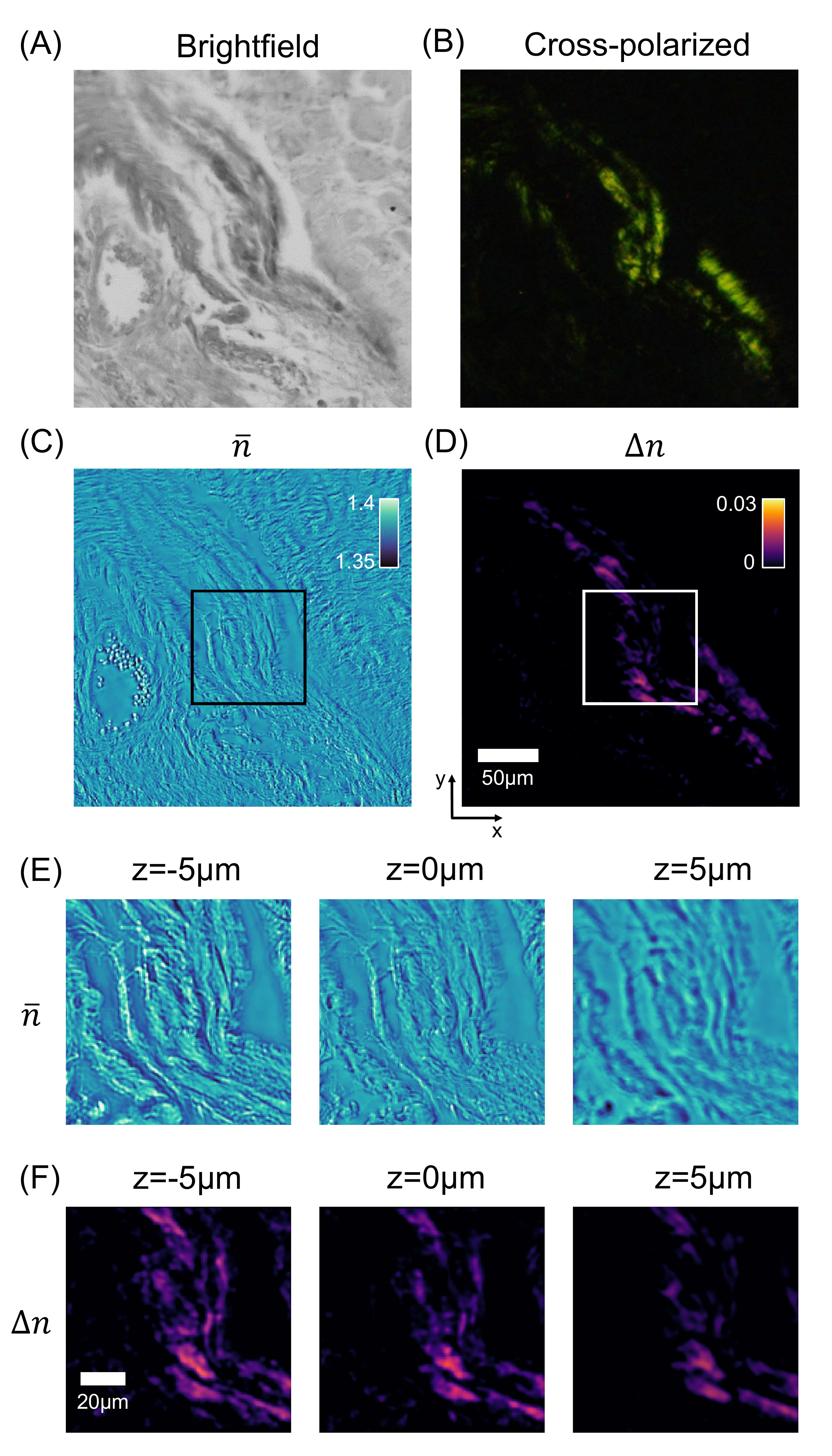}
\caption{Images of a heart tissue sample with cardiac amyloidosis. (A) displays a brightfield image. (B) shows a cross-polarized image taken with a color microscope. (C)-(D) present reconstructed refractive index and birefringence, along with zoom-ins of the boxed region at different depths depicted in (E)-(F). (C) and (E) share the same color bar, while (D) and (F) share another common color bar.}
\label{fig6::card}
\end{figure}
Finally, we apply our method to image a heart tissue sample that has cardiac amyloidosis. Cardiac amyloidosis is a lethal disease that affects more than 12,000 patients in the US alone with $<5\%$ 10-year survival rate~\cite{quock2018epidemiology}. In current practice, the biopsied tissue is first frozen and thinly sliced, then stained with a congo red-colored dye and inspected under a cross-polarized microscope. Figure~\ref{fig6::card}(A) shows the brightfield image. (B) shows the same region, imaged with a cross-polarized color microscope. The vibrant apple green color suggests mostly likely amyloid protein has built up inside the tissue sample. (C)-(D) depict the reconstructed refractive index and birefringence, while zoom-ins of the boxed region at different depths are shown in (E)-(F). Since the sample is thinly sliced, we do not observe noticeable structural changes in different layers. However, the structure of the birefringence reconstruction is correlated with the color-stained cross-polarized image, which could potentially be useful for rapid on-site inspections in the future.

\section{Discussion and Conclusion}
In this article, we introduce \tofu, a new non-scanning microscopy method that reconstructs volumetric permittivity metrics of samples based on computational illumination strategies to record polarized measurements and retrieve phase information. Using relatively low-NA objectives, we demonstrate that \tofu~can provide polarization-sensitive tomographic reconstruction for various calibration samples and biological specimens that are potentially useful for future scientific and clinical studies.

To ensure successful clinical translation, there are a few improvements that can be made in the next step. First, to increase the frame rate of \tofu, high flash-rate LEDs need to be deployed~\cite{reiser2008modular}. Novel sensitive camera sensors~\cite{tian2015computational,aidukas2019phase,yang2021quantized} will also play important roles in maintaining a good signal-to-noise ratio. In terms of the reconstruction algorithm, as a first demonstration, we use a gradient-based method implemented with auto-differentiation to optimize the loss function in Eq.~\ref{eq:loss_fuc_inverse}~\cite{paszke2019pytorch}. As the data fidelity term has a closed-form solution, variable-splitting methods can be very effective and allow incorporating advanced regularization that does not have implicit forms~\cite{sun2019regularized,zhou2020diffraction}. Moreover, due to spectral scanning, so far, \tofu\,has a lower frame rate compared to most holographic methods in the literature. To improve \tofu's throughput, in addition to using higher-speed LEDs and a more sensitive camera~\cite{li2019high}, adopting a multi-aperture approach appears to be another promising direction~\cite{wakefield2022cellular,harfouche2023imaging,thomson2022gigapixel,yang2023multi,yao2022increasing}. Finally, concerning modeling, to relieve the ill-posedness of the problem, we have approximated the permittivity tensor with its lateral components in this work. Albeit widely used~\cite{saba2021polarization,song2020ptychography,song2021large,dai2022quantitative,dai2021polarization,dai2020towards,xu2021imaging}, this simplification disregards out-of-plane anisotropy. Recent works have shown under uni-axial approximation, tomography of 3D polarization orientation can be retrieved~\cite{yeh2021upti}. Further investigation in this direction to extend the current method to extract valuable out-of-plane information is planned~\cite{yeh2021upti,yang2018polarized}.

\section*{Ethical approval}
Animal tissue sample studies were approved by the Friedrich-Alexander-University Erlangen-Nürnberg. All mice were maintained in the animal facility in a 12-h light–dark cycle with access to food and water. Human tissue sample studies were conducted with approval from the Duke University
Health System Institutional Review Board (DUHS IRB). The DUHS IRB determined that the following protocol meets the criteria for a declaration of exemption from further IRB review as
described in 45 CFR 46.101(b), 45 CFR 46.102 (f), or 45 CFR 46.102 (d), satisfies the Privacy
Rule as described in 45 CFR 164.512(i), and satisfies Food and Drug Administration regulations as described in 21 CFR 56.104, where applicable.

\section*{Funding}
The authors would also like to thank to a Duke-Coulter Translational Partnership and funding from a 3M Nontenured Faculty Award.

\section*{Conflict of interest}
S.X., X.D., L.K., J.N, C.G, and R.H. have submitted a patent application related to this work, assigned to Duke University.


\bibliographystyle{IEEEtran}
\bibliography{references}

\end{document}